\documentclass[pdflatex,sn-aps]{sn-jnl}% American Physical Society (APS) Reference Style
\usepackage{graphicx}%
\usepackage{multirow}%
\usepackage{amsmath,amssymb,amsfonts}%
\usepackage{amsthm}%
\usepackage{mathrsfs}%
\usepackage[title]{appendix}%
\usepackage{xcolor}%
\usepackage{textcomp}%
\usepackage{manyfoot}%
\usepackage{booktabs}%
\usepackage{algorithm}%
\usepackage{algorithmicx}%
\usepackage{algpseudocode}%
\usepackage{listings}%
\usepackage[table,xcdraw]{xcolor} % 支持颜色
\usepackage{graphicx}  % 插入图片
\usepackage{subcaption} % 支持子图
\usepackage{array} % 表格支持
\usepackage{booktabs} % 提供表格线样式
\usepackage{float} % 控制图表位置
\usepackage{lineno}
\theoremstyle{thmstyleone}%
\theoremstyle{thmstyletwo}%

\theoremstyle{thmstylethree}%

\begin{document}

\title[Article Title]{Simulation Study of Reconstruction of Cosmic Ray Extensive Air Shower with Thermal Neutron Detected by ENDA}


\author[1]{\fnm{Fan-Ping} \sur{Li}}
\author*[2,3]{\fnm{Xin-Hua} \sur{Ma}}\email{maxh@ihep.ac.cn}
\author*[1]{\fnm{Hu} \sur{Liu}}\email{huliu@swjtu.edu.cn}
\author[4,5]{\fnm{Tian-Lu} \sur{Chen}}
\author[6]{\fnm{Shu-Wang} \sur{Cui}}
\author[4,5]{\fnm{Danzengluobu} }
\author[4,5]{\fnm{Xin-Rui} \sur{Gao}} %\email{xrgao@ihep.ac.cn}
\author[2,3]{\fnm{Wei} \sur{Gao}}
\author[2,7]{\fnm{Wen-Chao} \sur{Gao}}
\author[4,5]{\fnm{Hai-Bing} \sur{Hu}}
\author[8]{\fnm{Denis} \sur{Kuleshov}}
\author[8]{\fnm{Kirill} \sur{Kurinov}}
\author[9]{\fnm{Bing-Bing} \sur{Li}}
\author[4,5]{\fnm{Mao-Yuan} \sur{Liu}}
\author[10]{\fnm{Ye} \sur{Liu}}
\author[8]{\fnm{Igor} \sur{Maliy}}
\author[6]{\fnm{Da-Yu} \sur{Peng}}
\author[6]{\fnm{Yao-Hui} \sur{Qi}}
\author[1]{\fnm{Jia-Bo} \sur{Qin}}
\author[4,5]{\fnm{Dong} \sur{Qu}}
\author[8,11]{\fnm{Oleg} \sur{Shchegolev}}
\author[8,11]{\fnm{Yuri} \sur{Stenkin}}
\author[4,5]{\fnm{Yu-Sen} \sur{Wang}}
\author[10]{\fnm{Tian-Shuang} \sur{Yang}}
\author[2,3]{\fnm{Li-Qiao} \sur{Yin}}
\author*[6]{\fnm{Hui-Qian}\sur{Zhang}}\email{hqzhang0331@163.com}
\author[6]{\fnm{Liang-Wei} \sur{Zhang}}
\author[6]{\fnm{Shi-Yuan} \sur{Zhang}} %\email{shyzhang@ihep.ac.cn}

\affil[1]{{School of Physical Science and Technology, Southwest Jiaotong University, Chengdu, Sichuan 610031, China}}
\affil[2]{{State Key Laboratory of Particle Astrophysics, Institute of High Energy Physics, Chinese Academy of Sciences, Beijing 100049, China}}
\affil[3]{{TIANFU Cosmic Ray Research Center, Chengdu, Sichuan 610000, China}}
\affil[4]{{Key Laboratory of Cosmic Rays, Ministry of Education, Xizang University, Lhasa, Xizang 850000, China}}
\affil[5]{{Science School, Xizang University, Lhasa, Xizang 850000, China}}
\affil[6]{{College of Physics, Hebei Normal University, Shijiazhuang 050024, China}}
\affil[7]{{University of Chinese Academy of Sciences, Beijing 100049, China}}
\affil[8]{{Institute for Nuclear Research of the Russian Academy of Sciences, Moscow 117312, Russia}}
\affil[9]{{College of Physical Science and Technology, Bohai University, Jinzhou, 121013, China}}
\affil[10]{{School of Management Science and Engineering, Hebei University of Economics and Business, Shijiazhuang, Hebei 050061, China}}
\affil[11]{{Moscow Institute of Physics and Technology, 141700 Moscow, Russia}}


\abstract{
In the investigation of the "knee" region within the cosmic ray energy spectrum, extensive air shower (EAS) reconstruction by using different EAS secondary particles are of paramount importance for energy determination and composition separation and finally measurement of energy spectrum of different cosmic ray composition. The Electron-Neutron Detector Array (ENDA) is designed to detect thermal neutrons generated in the ground by high energy hadrons which constitute the "backbone" of EAS and carry vital information regarding the primary cosmic rays. 
At present, ENDA-64, consisting of 64 detectors, is employed inside the Large High Altitude Air Shower Observatory (LHAASO). This work simulates the response of ENDA and investigates the lateral distributions of thermal neutrons. A comparison of the NKG and double-exponential functions for describing this distribution using MC simulation data shows that the double-exponential function is more accurate. The mean difference between fitted and measured thermal neutron density is within $\sim$ 6\%. Meanwhile, the thermal neutron density at 10 m from the shower axis, $\rho_n$, depends mainly on the sum of all hadrons energy or the number of high-energy hadrons. The relationship between $\rho_n$ and EAS hadron energy or the number of high-energy hadrons can be well fitted by a linear function on a log scale. This work is fundamental for energy determination and composition separation with hybrid detection  by using ENDA in the near future. \\\indent
}



\maketitle

\section{Introduction}\label{Introduction}
When one high-energy cosmic ray nucleus enter Earth's atmosphere, it initiates extensive air shower (EAS) through hadronic and electromagnetic cascades interactions with atmospheric nuclei. EAS includes several types of secondary particles: electromagnetic components which then produce atmospheric Cherenkov lights, fluorescence lights and radio waves; hadronic components which occupy the shower core; muons and neutrinos which are mainly produced by decay of low energy hadrons. Hybrid detection of secondary particles in EAS provides key insights into cosmic ray composition and energy spectrum. The lateral distribution of secondary particles in EAS is crucial for reconstructing EAS direction, core position, shower size and shower age parameters which are the foundation of energy determination and composition separation of cosmic rays.

The Large High Altitude Air Shower Observatory (LHAASO)\cite{lhaaso_chapter1} is located at Mt. Haizi (4410 m a.s.l., 600 g/cm$^2$, 29$^{\circ}$ 21' 27.56" N, 100$^{\circ}$ 08' 19.66" E) in Daocheng, Sichuan province, P.R. China. The observatory comprises a 1.3 $\mathrm{km}^2$ array (KM2A) of electromagnetic (ED) and muon detectors (MDs) for $\gamma$-ray and cosmic-ray studies, a water Cherenkov detector array (WCDA) for TeV $\gamma$-ray observations, and 18 wide-field air Cherenkov/fluorescence telescopes (WFCTA) dedicated to cosmic-ray measurements. It records the electromagnetic component, muonic component, and Cherenkov light generated by extensive air showers (EAS). LHAASO reconstruction details can be found in reference~\cite{KM2ACrab,WFCTARec,WCDARec}.
. 
A newly proposed electron-neutron detector array (ENDA) \cite{lhaaso_chapter1}\cite{2023PAN....86.1056P}\cite{2025PAN....88.1147P}, which detects thermal neutrons from secondary hadrons on the ground, has been deployed inside LHAASO covering 1,000 m$^2$ in phase one at present and will be extended to 10,000 m$^2$ in phase two in the future. Obviously, combination between LHAASO and ENDA fulfills an unique experiment on detection of all secondary particles in EAS so as to obtain more accurate measurement of cosmic ray composition and energy spectrum.

%++++++++++++++++++++++++++++++++++++++++++++++++++++
\section{Experimental Setup}
\label{experiment}

EAS secondary hadrons, which constitute the EAS skeleton, carry important information for multi-parameter correlation studies. When hadrons interact with ground matter (e.g., soil or building materials), they generate evaporation neutrons. These neutrons gradually lose energy through elastic scattering until reaching thermal equilibrium, defined as thermal neutrons. Thermal neutrons can provide important information about parent hadrons and, consequently, about the energy and composition of cosmic rays~\cite{Stenkin2002MPLA,STENKIN2009293,Stenkin_2013,Stenkin2007,STENKIN2008326}. ENDA is designed to detect thermal neutrons as well as electromagnetic components near EAS cores. Currently, ENDA comprises 64 electron-neutron detectors (EN-detectors) organized into four clusters. Each cluster is composed of 16 detectors with 4.5~m spacing.  One EN-detector with sensitive area of 0.35~m$^2$ uses ZnS(Ag) scintillator alloyed with B$_2$O$_3$ with effective thickness of 50~mg/cm$^2$). The scintillator layer is coupled to a 4-inch PMT inside a polyethylene tank. The detector distinguishes early and fast charged-particle signals from delayed slow thermal-neutron capture signals in one EAS. Neutron sampling time window is 5 ms. ENDA-64 will give results of measurement of cosmic ray energy spectrum around the knee region of light component (proton and helium). After it, the array is planned to be extended to the ENDA-400 composed by 400 EN-detectors to measure the  knee region of heavy components (Fe). Details of the array can be found in ~\cite{lhaaso_chapter1,2023PAN....86.1056P,2025PAN....88.1147P,2025RDTM....9..600Z}.

%++++++++++++++++++++++++++++++++++++++++++++++++++++
\section{Simulation Setup}
\hspace{2em}The EAS development initiated by primary cosmic rays was simulated using CORSIKA (Cosmic Ray Simulations for Kascade) version {7.7410}\cite{corsika}. QGSJET-II is utilized for high-energy hadronic interactions, and FLUKA is employed for low-energy hadronic interactions. For CORSIKA events, we simulate KM2A and ENDA detector responses with identical shower core positions enforced across both arrays. The secondary particles reaching KM2A were processed by a dedicated detector simulation program (named G4KM2A~\cite{2014APh....54...86C,KM2ASim, KM2ASim2,2024RDTM....8.1437C}) in the framework of Geant4 toolkit version 10.4~\cite{1610988}. The reliability of the detector simulation was verified via the KM2A array data, detailed comparison between data and MC simulation can be found in reference~\cite{KM2ASim2,2024RDTM....8.1437C}. \\\indent
Secondary particles arriving at the ENDA are processed via a dedicated Geant4 v10.4 simulation framework, whose configuration parameters rigorously derived from experimentally measured values: structural parameters of the detector array, material properties of components, and electronics response characteristics. The selected hadronic interaction model of ENDA simulation is QGSP\_BIC\_HP, a combination of the Quark Gluon String Model (QGS, $>\sim$20 GeV), the Binary Cascade Model (BIC, $<\sim$10 GeV), and the High Precision Neutron Model (HP, $<$ 20 MeV). ~Details of simulation methods and software structure were described in reference~\cite{2024FrPhy..1944200L} except that now number of the ENDA detectors is extended from 16 to 64. The negative correlation between the thermal neutron counting rate and soil moisture was obtained by employing 5 soil moisture meters inside ENDA~\cite{Yang_2023}, and thermal neutrons detected by ENDA will be corrected with the correlation parameters. Therefore at present, in simulation, soil is set dry as the baseline.  

In this work, the whole primary energy range is set from 100 TeV to 10 PeV. The zenith angle range is set from 0$^\circ$ to 40$^\circ$, the azimuthal angle range is set from 0$^\circ$ to 360$^\circ$, and the observation level is set at 4400 m above sea level.  
The low energy cut-off of the secondary hadrons is set as 0.05 GeV. 

In the simulation, the shower cores are uniformly dropped within a 48 m radius circle centered on ENDA. the shower direction and  core position are obtained by LHAASO-KM2A~\cite{KM2ACrab}.

%++++++++++++++++++++++++++++++++++++++++++++++++++++
\section{Reconstruction Methods and Results}

%% 插入表格
%\begin{table}
%\centering
%\renewcommand{\arraystretch}{1.5} % 调整行高
%\caption{Fitting parameters of electron $\eta$ distribution of the five components. }
%\begin{tabular}{|c|c|c|c|c|c|}
%\hline

%% 插入表格
%\begin{table}
%\centering
%\renewcommand{\arraystretch}{1.5} % 调整行高
%\caption{Fitting parameters of muon $\eta$ distribution of the five components. }
%\begin{tabular}{|c|c|c|c|c|c|}
%\hline
 
Thermal neutrons are primarily produced by moderating evaporation neutrons from ground-level hadronic interactions. Their multiplicity exceeds that of their parent hadrons by up to two orders of magnitude\cite{Stenkin_2013}. The lateral distribution of secondary particles in extensive air showers is essential for ground-based cosmic-ray observations, forming the basis for reconstructing the energy and composition of primary particles. \\\indent
The Nishimura-Kamata-Greisen(NKG hereafter) function typically describes the lateral distribution of the electromagnetic component. For hadrons, early studies focused on the region within 10 meters of the shower axis, initially using a power-law parameterization\cite{Cocconi1961}. Later, the Maket-ANI collaboration reported consistency with an exponential form within 5 meters of the shower axis\cite{maket}. KASCADE's large hadronic calorimeter extended the analysis of the hadronic lateral distribution to larger distances, finding the NKG function provided the optimal fit\cite{kascade}. The scale radius $R_{m}$ $\approx$10 m for the hadronic component is much smaller than that for the electromagnetic component. The PRISMA-32 array parameterized thermal neutrons using a double exponential function \cite{doubleexp}. This shape comprises a steep component from prompt neutrons near the shower axis and a flatter component from delayed neutrons produced in ground-level interactions farther out or in air at higher levels. This study compares the NKG function (used for hadrons in KASCADE) and the double exponential function (Equation \ref{eq:double_exp}; $N_n$: integrated neutron shower size, $f$: near component fraction), used for thermal neutrons in PRISMA-32, to describe thermal neutron lateral distributions for ENDA. \\\indent
\begin{equation}
\label{eq:double_exp}
\rho(r) = N_{n}\left( \frac{f}{2\pi r_1^2}\exp(-\frac{r}{r_1}) + \frac{1-f}{2\pi r_2^2}\exp(-\frac{r}{r_2}) \right)
\end{equation}
\indent

\begin{figure*}[htbp]
\centering
\begin{minipage}[t]{1.\linewidth}
\includegraphics[width=0.48\linewidth]{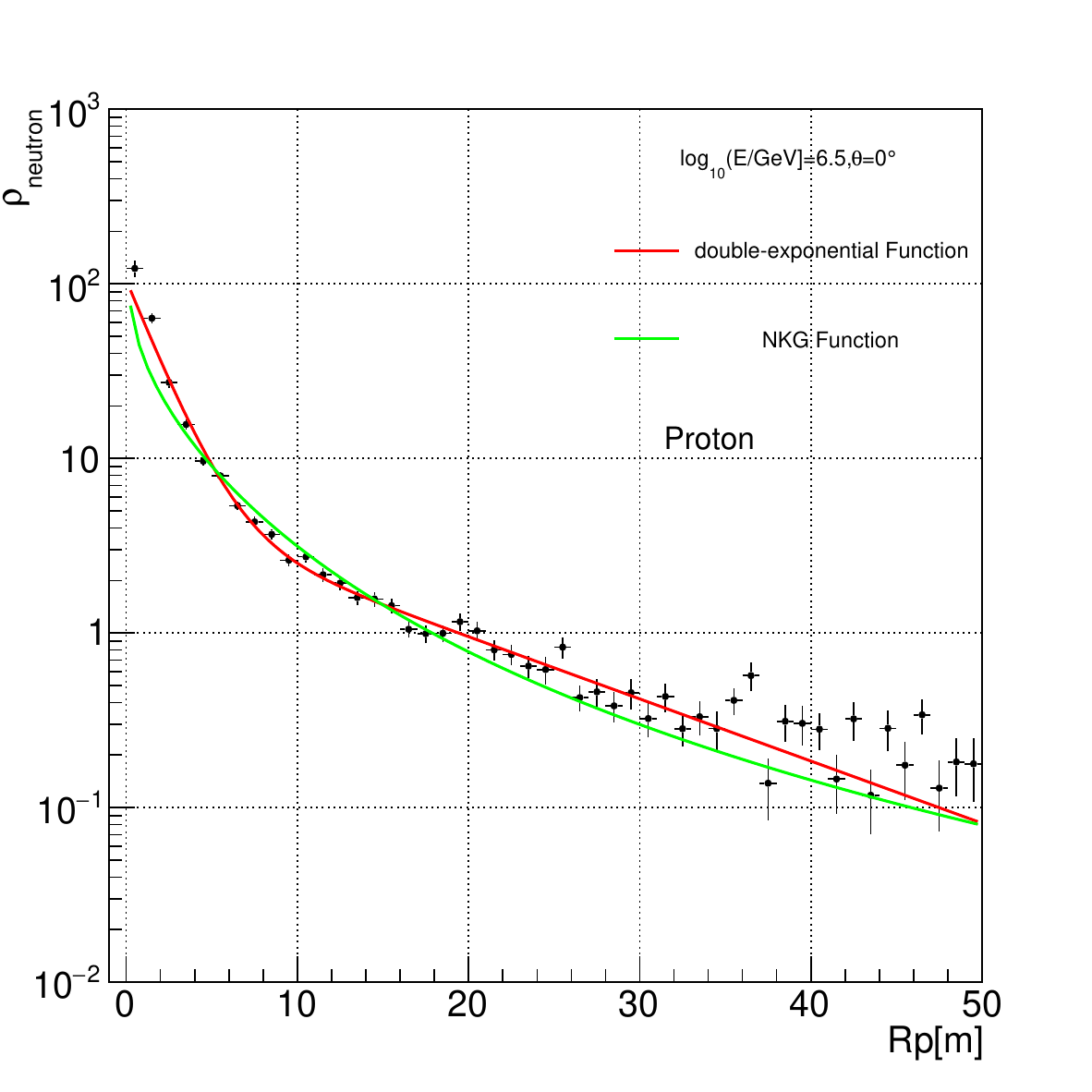}
\hspace{0cm}
\includegraphics[width=0.48\linewidth]{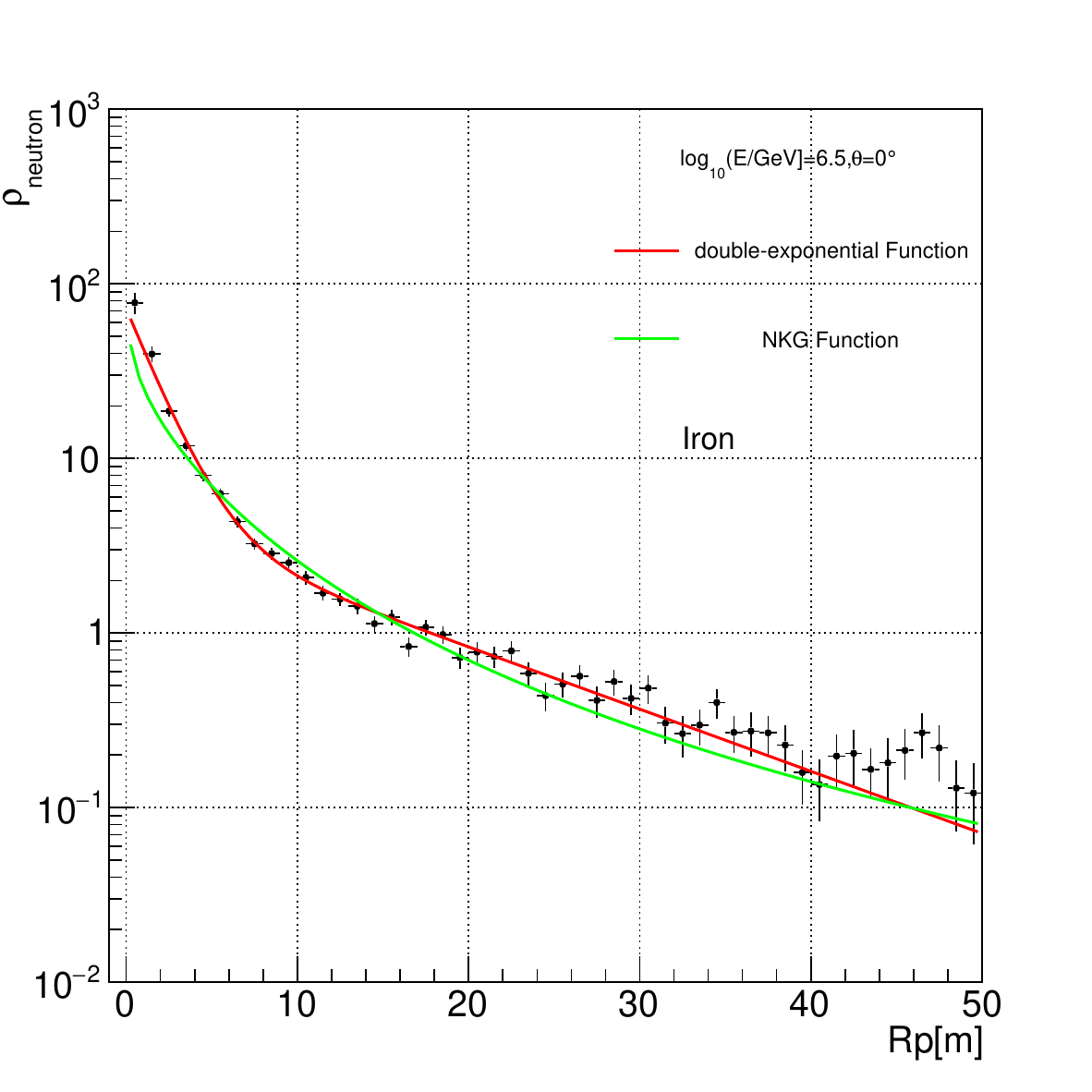}
\caption{Comparison of fitting functions for the thermal neutron average lateral distribution of proton (left) and iron (right) at $\log_{10}(E/\text{GeV}) = {6.5}$ with $\theta = 0^\circ$. The NKG function (green curve) and double-exponential function (red curve) are shown with $R_M = 10$ m and $r_1 = 1.77 m$, $r_2 = 12.20 m$ respectively. }
\label{fig:fitting_comparison}
\end{minipage}
\end{figure*}
\begin{figure*}[htbp]
\centering
\begin{minipage}[t]{1.\linewidth}
\includegraphics[width=0.48\linewidth]{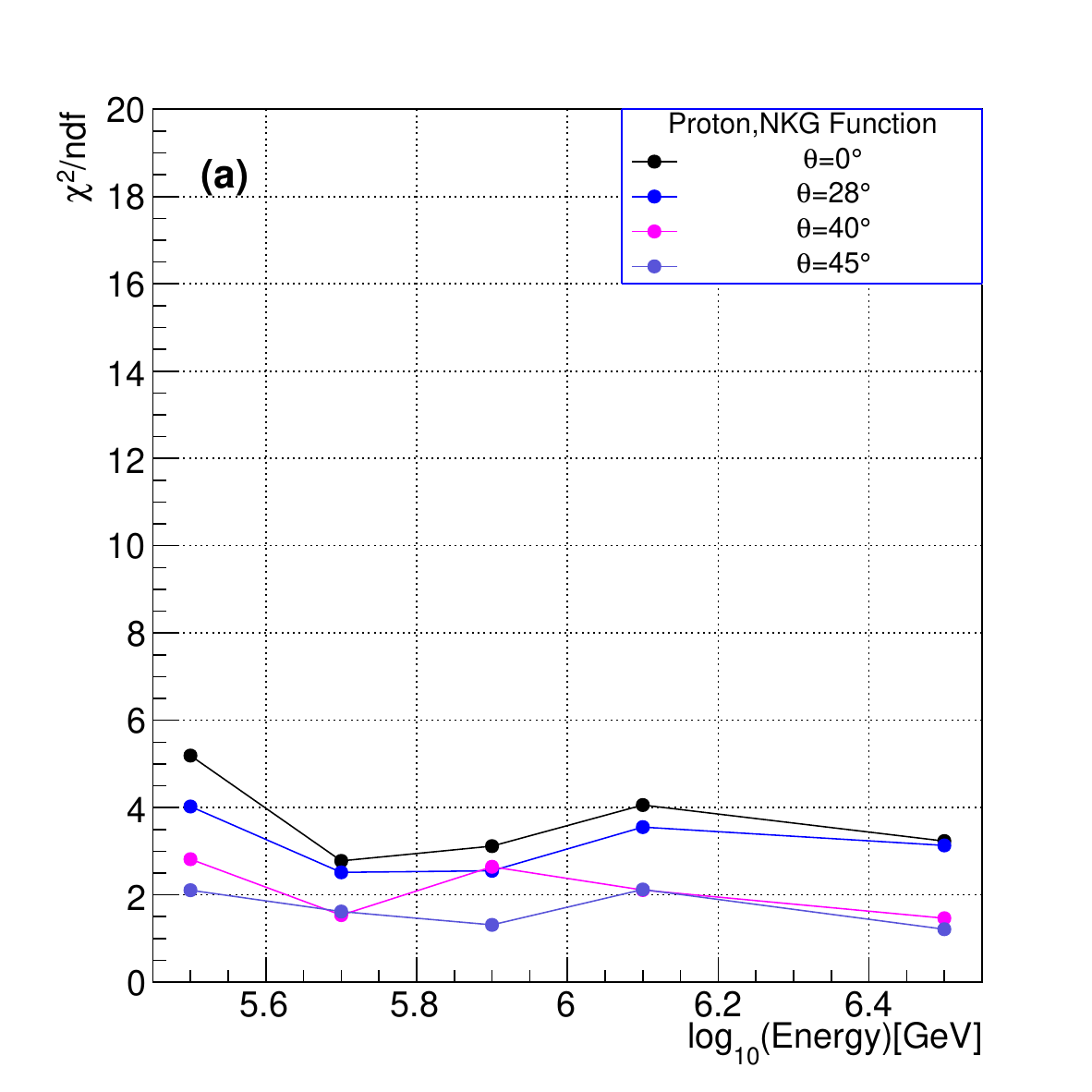}
\hspace{0cm}
\includegraphics[width=0.48\linewidth]{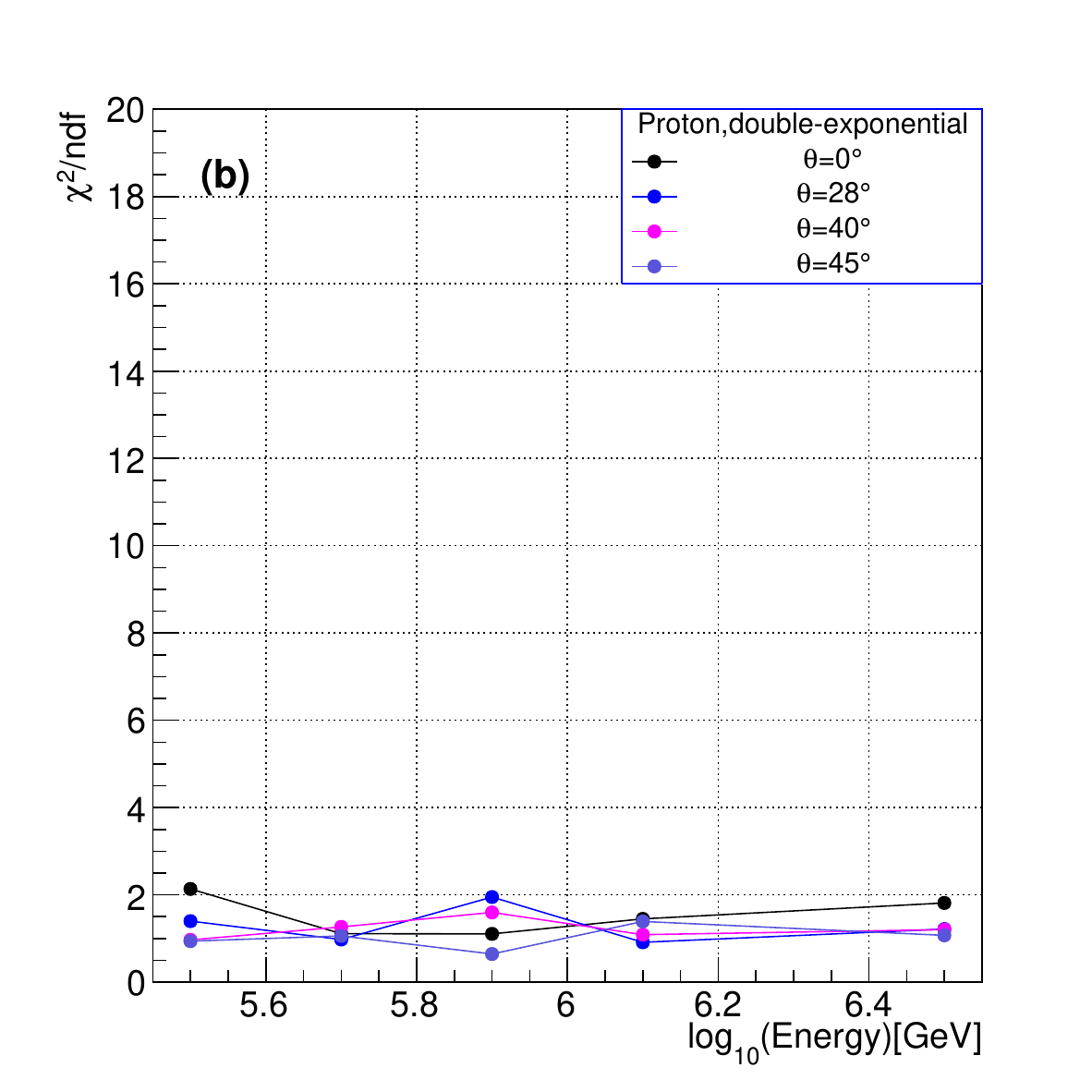}
\hspace{0cm}
\includegraphics[width=0.48\linewidth]{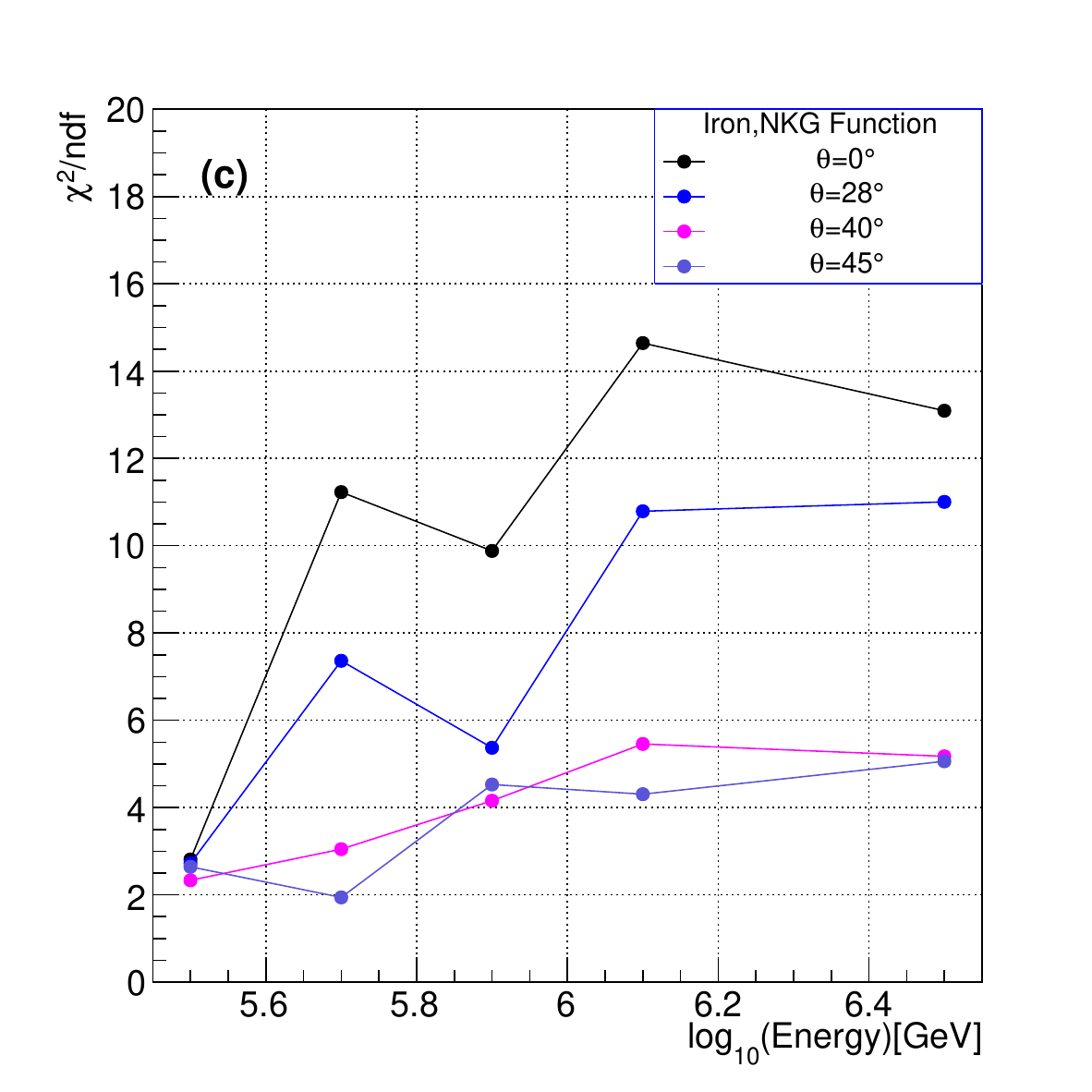}
\hspace{0cm}
\includegraphics[width=0.48\linewidth]{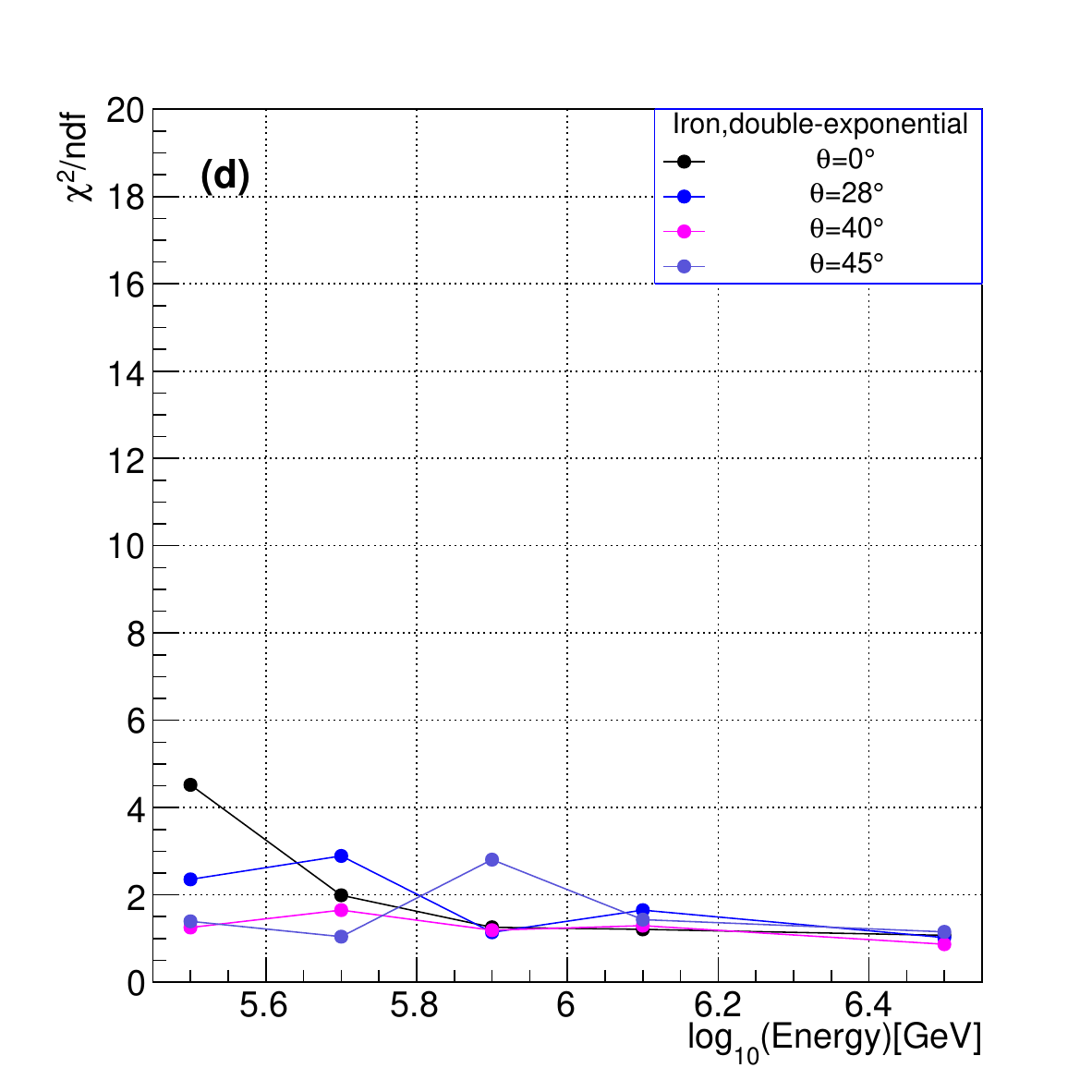}
\caption{The $\chi^2/\mathrm{ndf}$ values from NKG (panels a, c) and double-exponential (panels b, d) fits plotted against primary energy. Results are shown for proton samples (a, b) and iron samples (c, d) across multiple zenith angles.}
\label{fig:chi2ndf}
\end{minipage}
\end{figure*}
Fig.~\ref{fig:fitting_comparison} compares NKG and double-exponential fits to the thermal neutron average lateral distribution for vertical showers at $E = 10^{6.5}$ GeV, with proton (left) and iron (right) primaries. The derived parameters for the double-exponential function are $r_1 = 1.77\pm0.03$ m (near component) and $r_2 = 12.20\pm0.13$ m (far component). Fig.~\ref{fig:chi2ndf} presents $\chi^2/\mathrm{ndf}$ comparisons of NKG and double-exponential fits for samples spanning different primary compositions, energies, and zenith angles. The parameters $r_1$ and $r_2$ of double-exponential function are fixed for all samples for the reasons outlined below. As shown, the double-exponential function fits to the thermal neutron lateral distribution have $\chi^2/\mathrm{ndf}$ closer to one, and is better than the NKG function. \\\indent
\begin{figure*}[htbp]
\centering
\begin{minipage}[t]{1.\linewidth}
\includegraphics[width=0.32\linewidth]{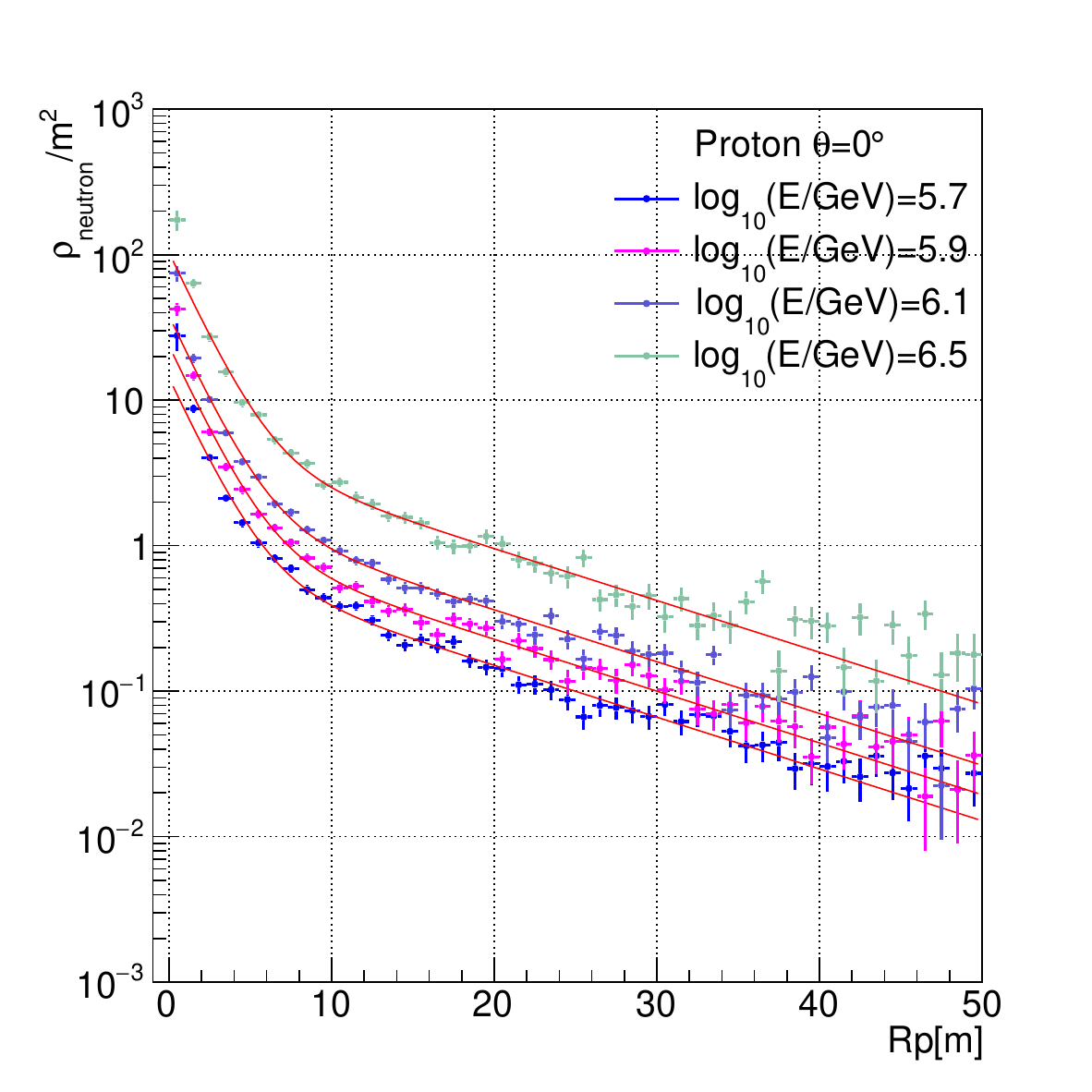}
\hspace{0cm}
\includegraphics[width=0.32\linewidth]{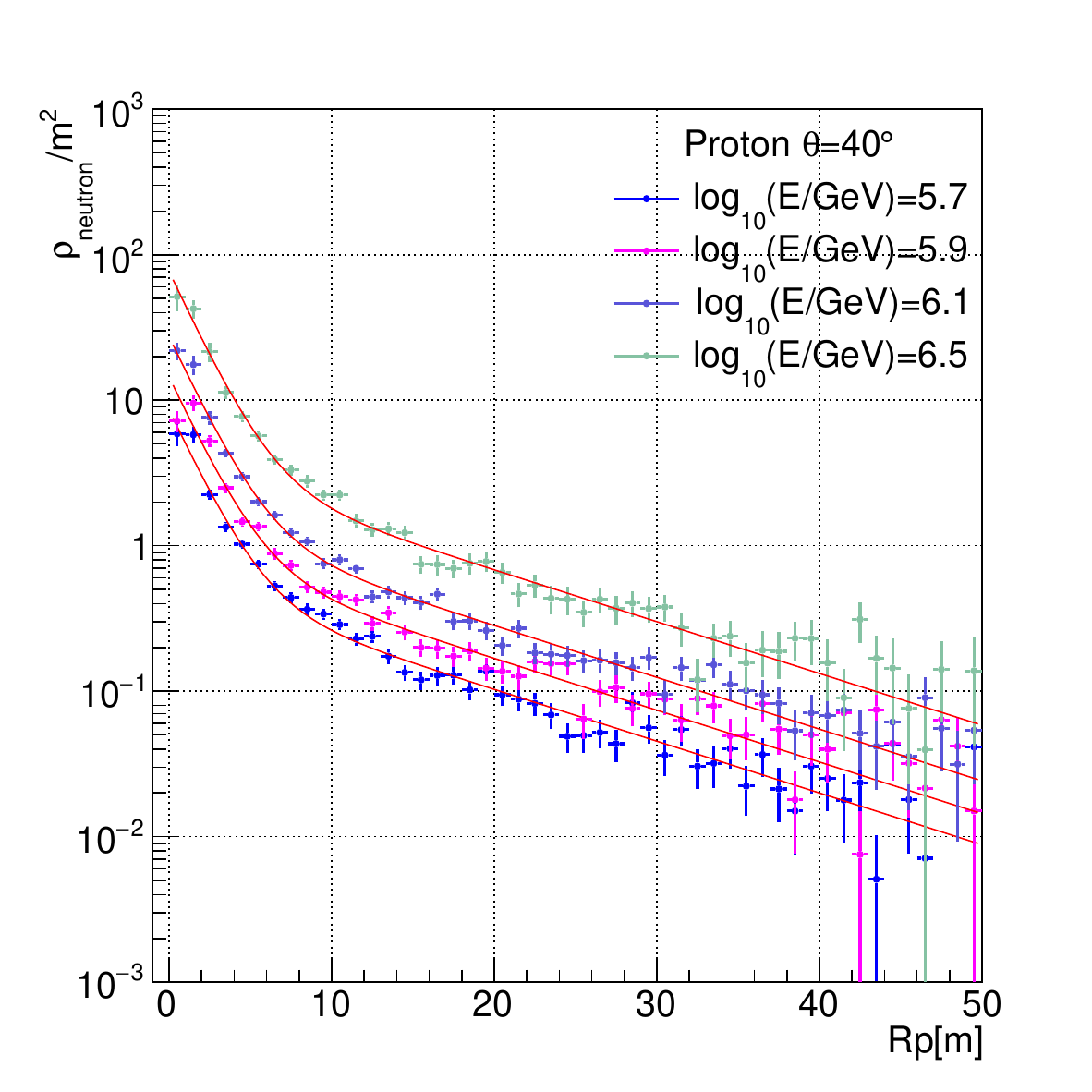}
\hspace{0cm}
\includegraphics[width=0.32\linewidth]{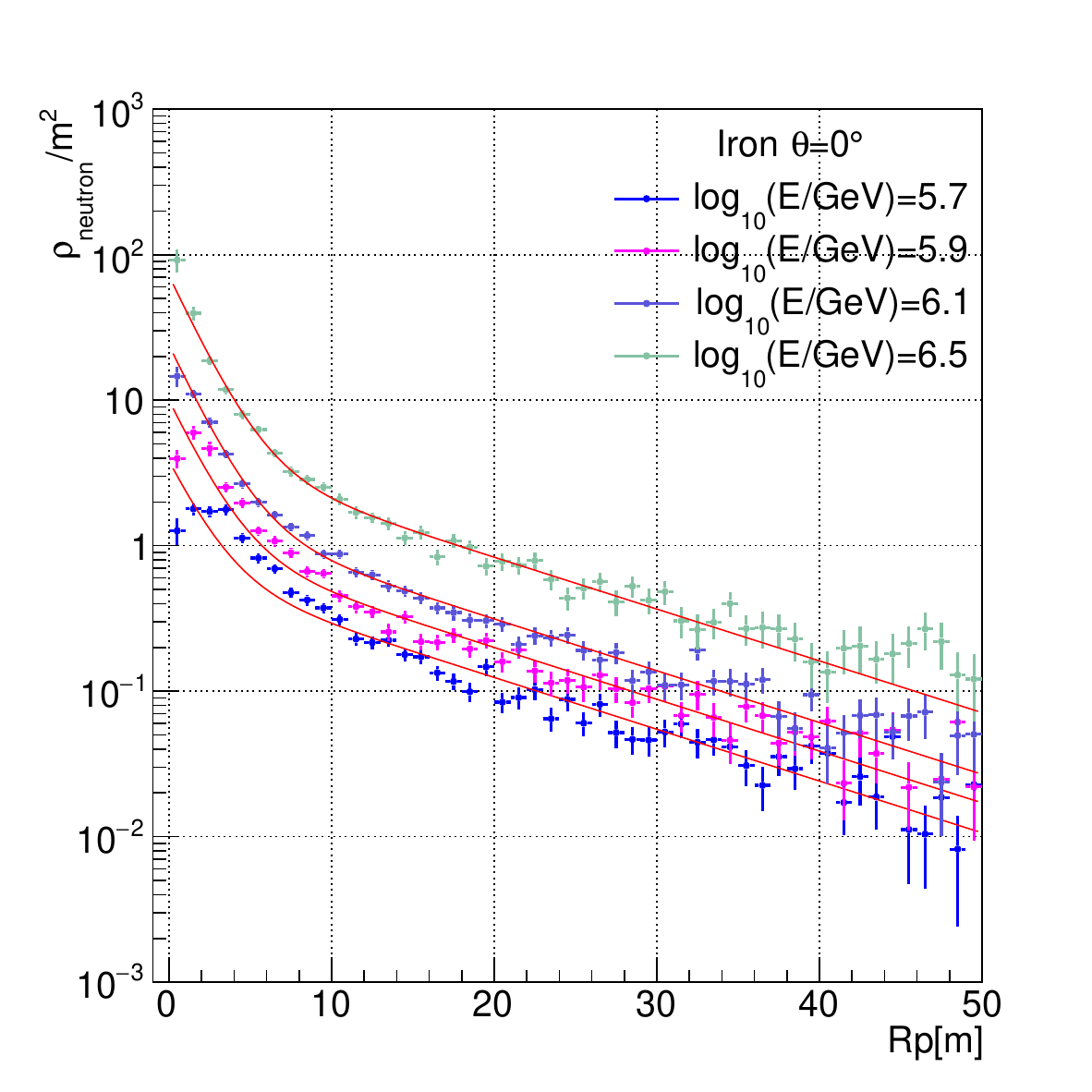}
\caption{Average lateral distributions of thermal neutrons for: vertical-incident protons (left), protons at $\theta = 40^\circ$ (middle), and vertical-incident iron (right) at $\log_{10}(E/\text{GeV}) = {{5.7,5.9,6.1,6.5}}$. The lines are double-exponential fit to the data points.}
\label{fig:lateral_profile}
\end{minipage}
\end{figure*}

To reduce free parameters given the limited number of fired EN-detectors, we try to fix the parameters $r_1 = 1.77$ m and $r_2 = 12.20$ m, values extracted from proton shower samples. Only the thermal neutron shower size $N_n$ and near-component fraction $f$ in Eq.~\ref{eq:double_exp} are left free to vary. Fig.~\ref{fig:chi2ndf}b and~\ref{fig:chi2ndf}d displays the $\chi^2/\mathrm{ndf}$ values from double-exponential fits across varying primary compositions, energies, and zenith angles. The fits yield $\chi^2/\mathrm{ndf}$ less than 3 with fixed $r_1$ and $r_2$, except for vertical iron samples below 400 $\mathrm{TeV}$. These results indicate that $r_1$ and $r_2$ exhibit no strong dependence on energy, primary composition, and zenith angle. Fig.~\ref{fig:lateral_profile} shows the average lateral distributions for proton and iron showers, fitted with the double-exponential function using the fixed $r_1$ and $r_2$ across several energies and zenith angles.\\\indent

\begin{figure*}[htbp]
\centering
\includegraphics[width=0.48\textwidth]{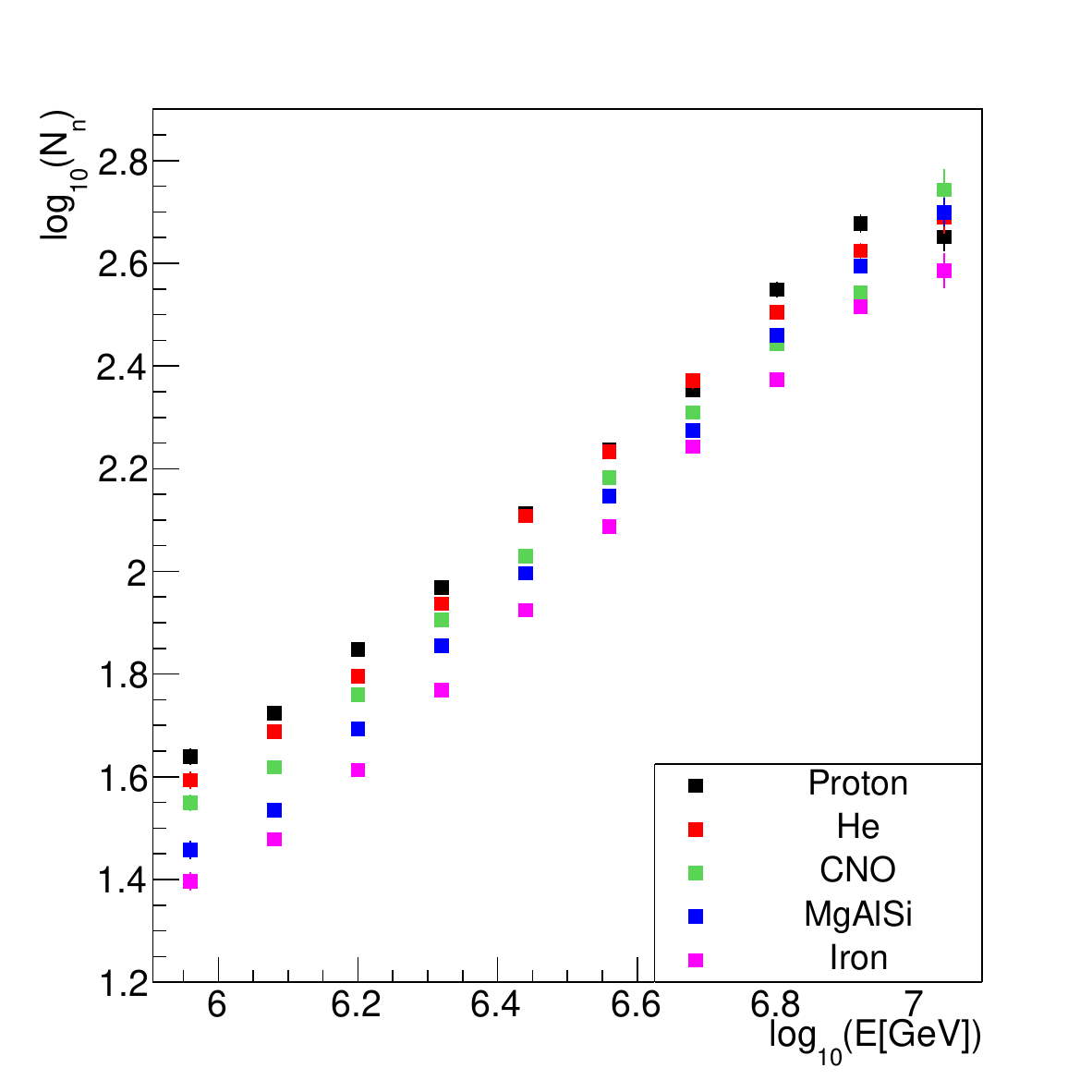}

\caption{Relation between average thermal neutron shower size $N_n$ and primary energy $E$ for different components.}

\label{fig:ehoE}
\end{figure*}
\begin{equation}
\label{eq:likelihood}
L(N_n,f)=\prod\limits_{i=1}^n \frac{(\rho(r_i)S)^{N_i}}{N_i!}e^{-\rho(r_i)S}
\end{equation}
\\\indent
Since the thermal neutron counts recorded by each EN-detector are low, per-detector neutron counts follow a Poisson rather than Gaussian distribution, maximum likelihood fitting method is better suited for these low-statistics regimes. The likelihood is defined in Eq.~\ref{eq:likelihood}. Here, n denotes the number of fully functional ENDA detectors, $N_i$ the thermal neutron count measured by the i-th detector, $\rho(r_i)$ the expected thermal neutron density derived from the shower lateral distribution in Eq.~\ref{eq:double_exp}, where $N_n$ and $f$ are treated as free parameters; and S the detection area of a single EN-detector. We use the ROOT TMinuit package to maximize the likelihood function~\cite{root-tminuit}. This fitting procedure outputs the event-level parameters $N_n$ and $f$ from Eq.~\ref{eq:double_exp}, which together describe the lateral distribution of thermal neutrons. Fig.~\ref{fig:ehoE} shows the correlations of average thermal neutron shower size $N_{n}$ with primary energy ($E$) for different composition. The correlations between $N_n$ and primary energy are distinct for different primary nuclei. \\\indent

\begin{figure*}[htbp]
\centering
\begin{minipage}[t]{1.\linewidth}
\includegraphics[width=0.48\linewidth]{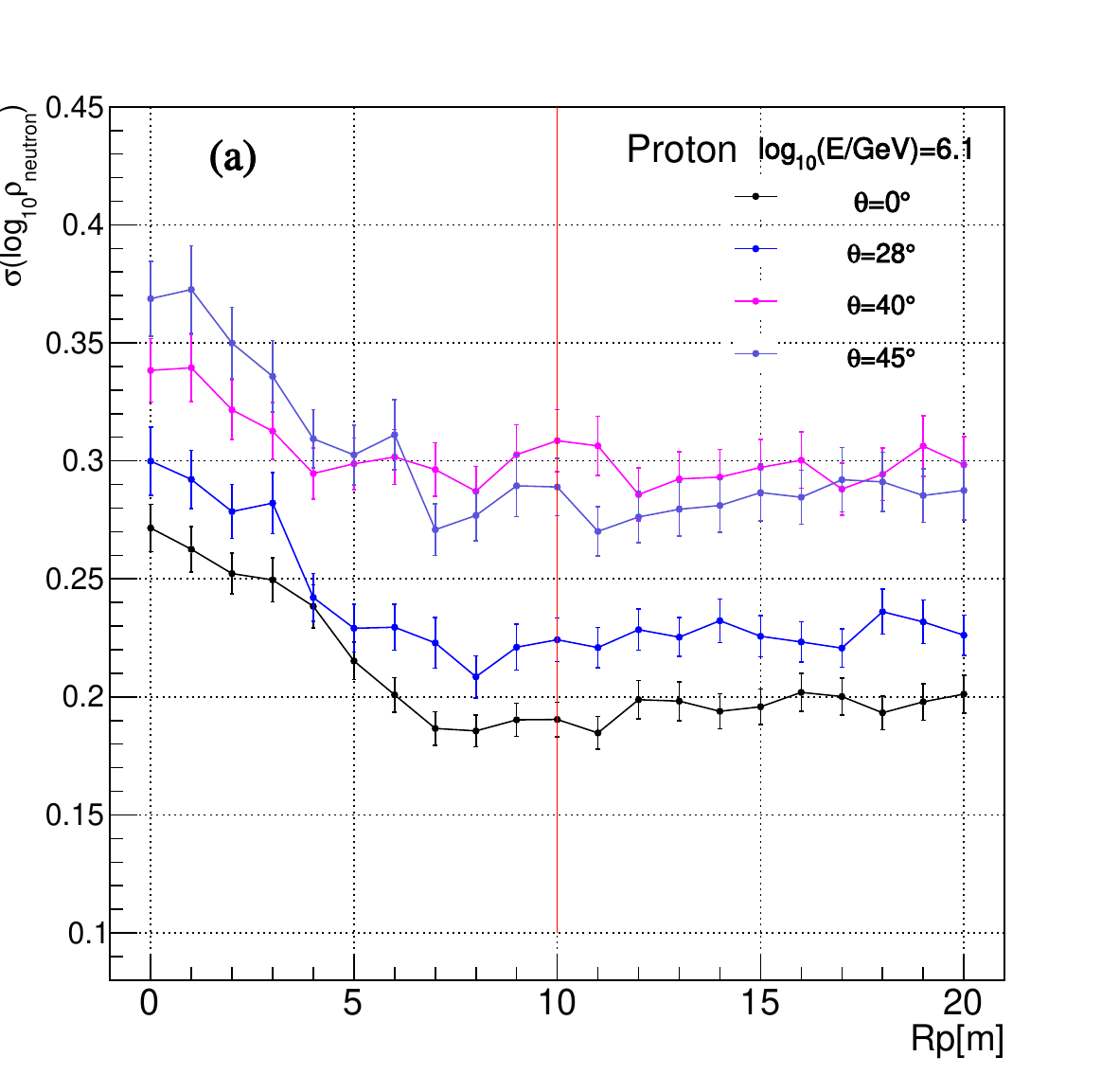}
\hspace{0cm}
\includegraphics[width=0.48\linewidth]{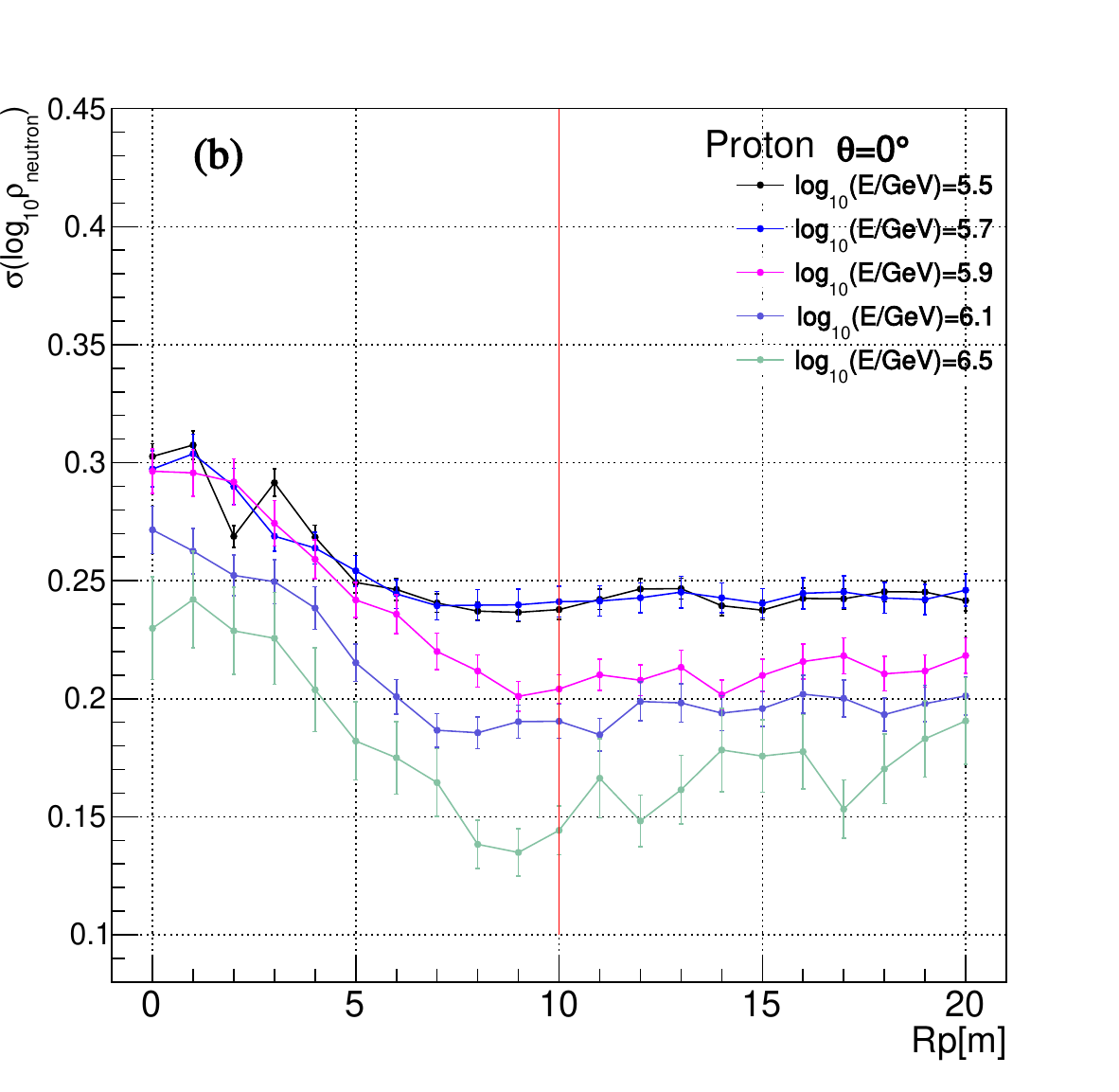}
\hspace{0cm}
\includegraphics[width=0.48\linewidth]{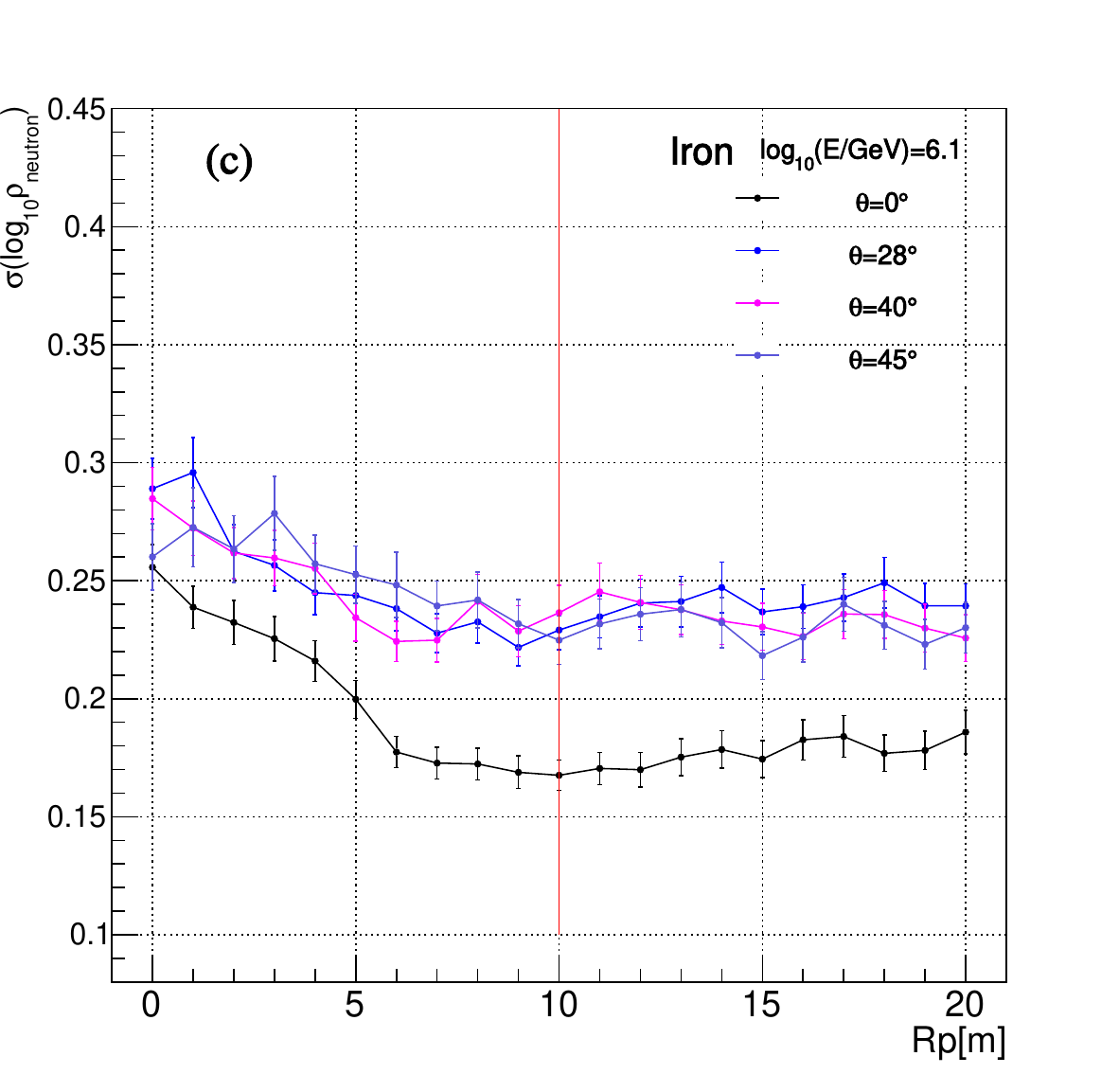}
\hspace{0cm}
\includegraphics[width=0.48\linewidth]{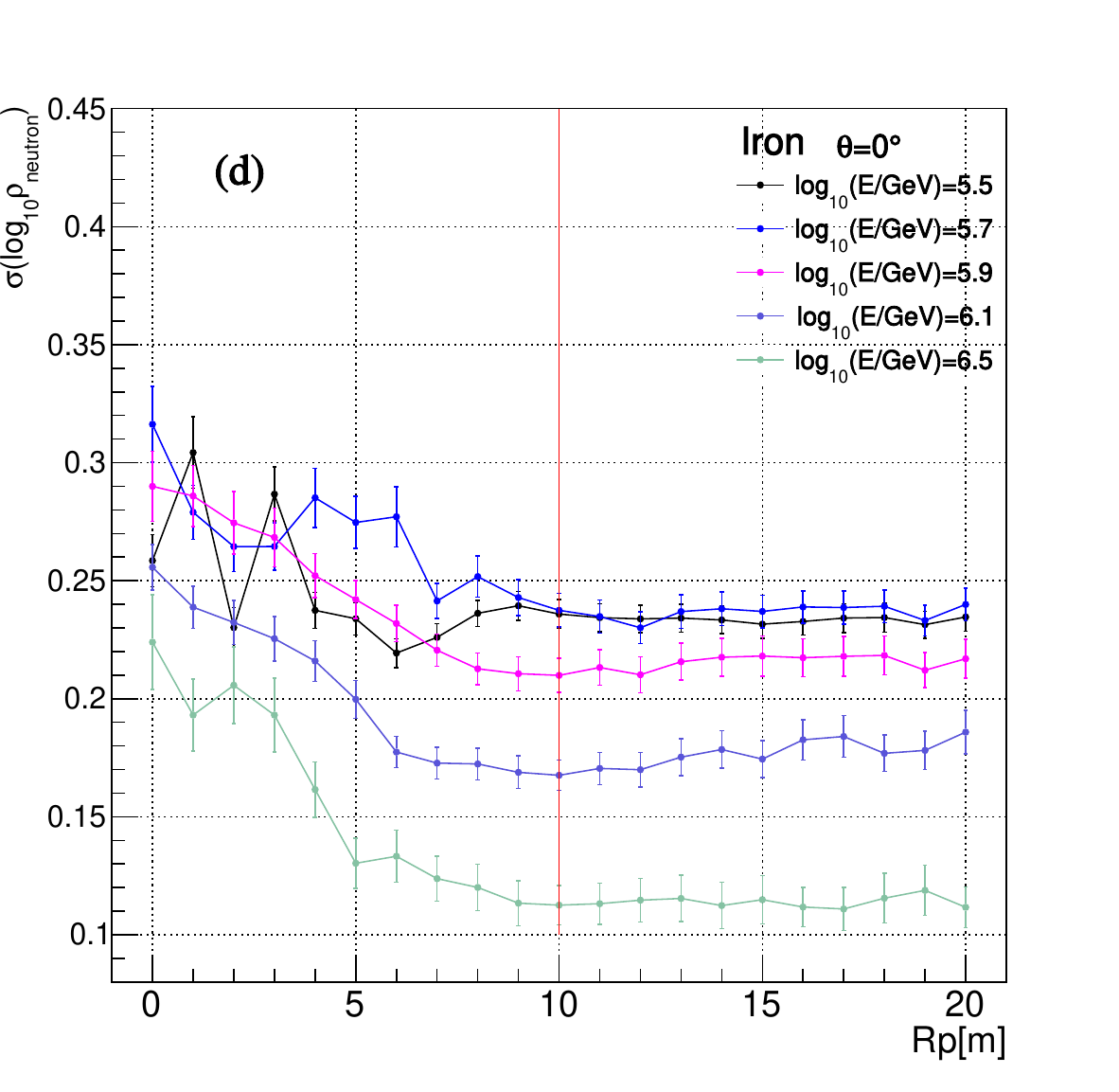}
\caption{Thermal neutron density resolution plotted against shower impact distance $R_p$. Results for proton samples (panels a, b) and iron samples (panels c, d) are shown across different energies (b, d) and zenith angles (a, c). Vertical red lines mark the selected thermal neutron density at $10\,\mathrm{m}$.}
\label{fig:rho-r}
\end{minipage}
\end{figure*}
Particle number densities derived from fitted lateral distribution are widely adopted for energy and composition reconstruction in ground-based experiments~\cite{Asg,Icetop,spos}. Here, we extract thermal neutron number density from the fitted lateral distribution. Fig.~\ref{fig:rho-r} shows the thermal neutron density resolution as a function of shower impact distance $R_p$ for various primary compositions, energies, and zenith angles. This resolution is defined as the width $\sigma$ obtained from Gaussian fits to thermal neutron density distributions evaluated at fixed shower impact distance. The $R_p$ minimizing $\sigma$ varies across conditions, yet the thermal neutron density at 10 m (the vertical red line in Fig.~\ref{fig:rho-r}) yields good resolution for all primary compositions, energies, and zenith angles. Additionally, three observables extracted from ENDA—the thermal neutron density at 10 m from the shower axis ($\rho_{n}$), the thermal neutron shower size $N_{n}$, and the sum of thermal neutrons from all triggered detectors ($N_{sum}$)—are compared in reference~\cite{fanping_icrc}. Among the three variables, the thermal neutron density $\rho_{n}$ has the minimum shower-to-shower fluctuation.\\\indent
\begin{figure}[htbp]
    \centering
    \includegraphics[width=0.5\textwidth]{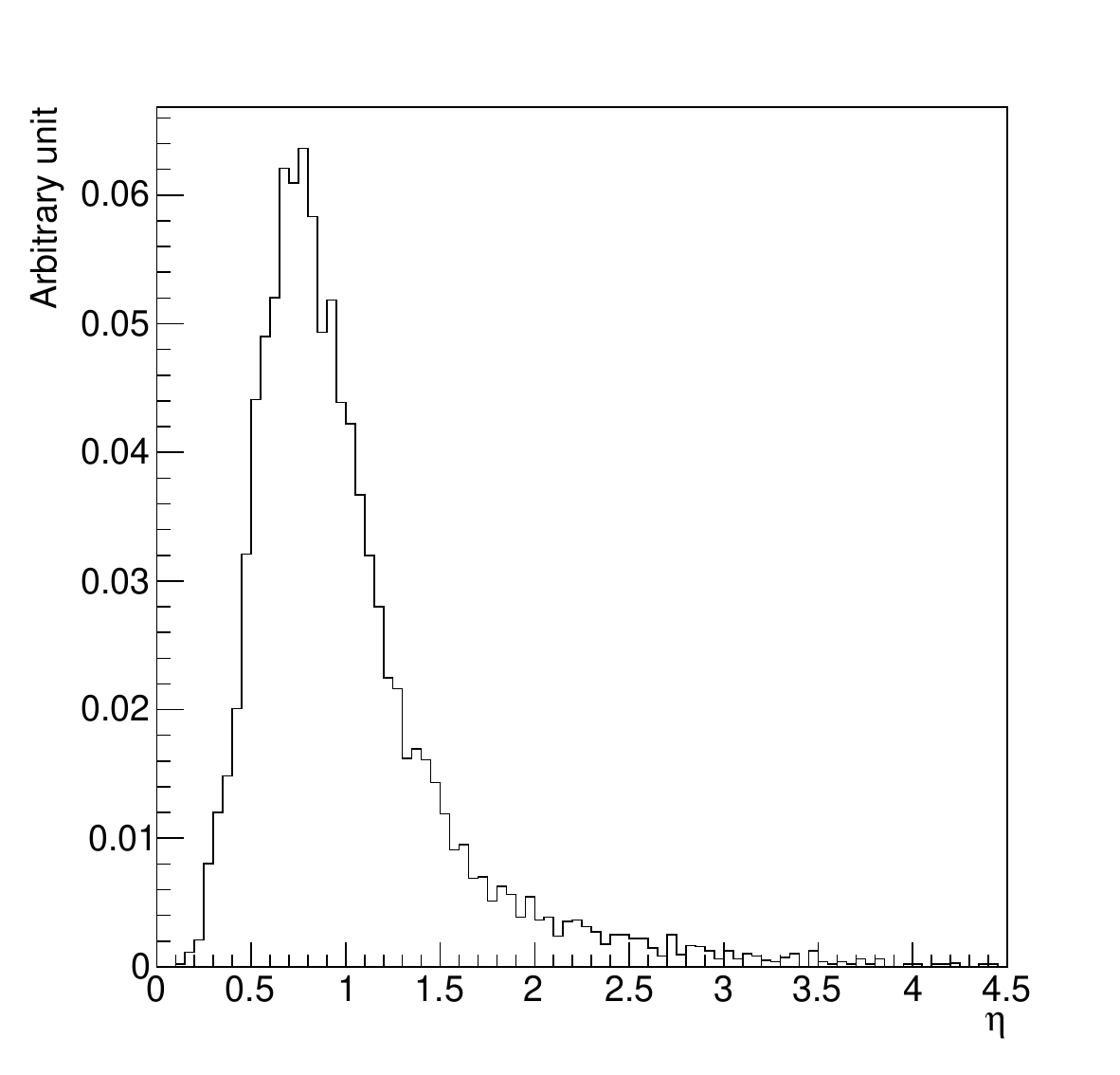}
    \caption{Thermal neutron $\eta$ distribution of proton sample. $\eta$ is defined as the ratio of fitted to measured thermal neutron densities at $10\,\mathrm{m}$, as formulated in Eq.~\ref{eq:eta}.}
    \label{fig:proton-etanu}
\end{figure}

\begin{figure}[htbp]
    \centering
    \includegraphics[width=0.5\textwidth]{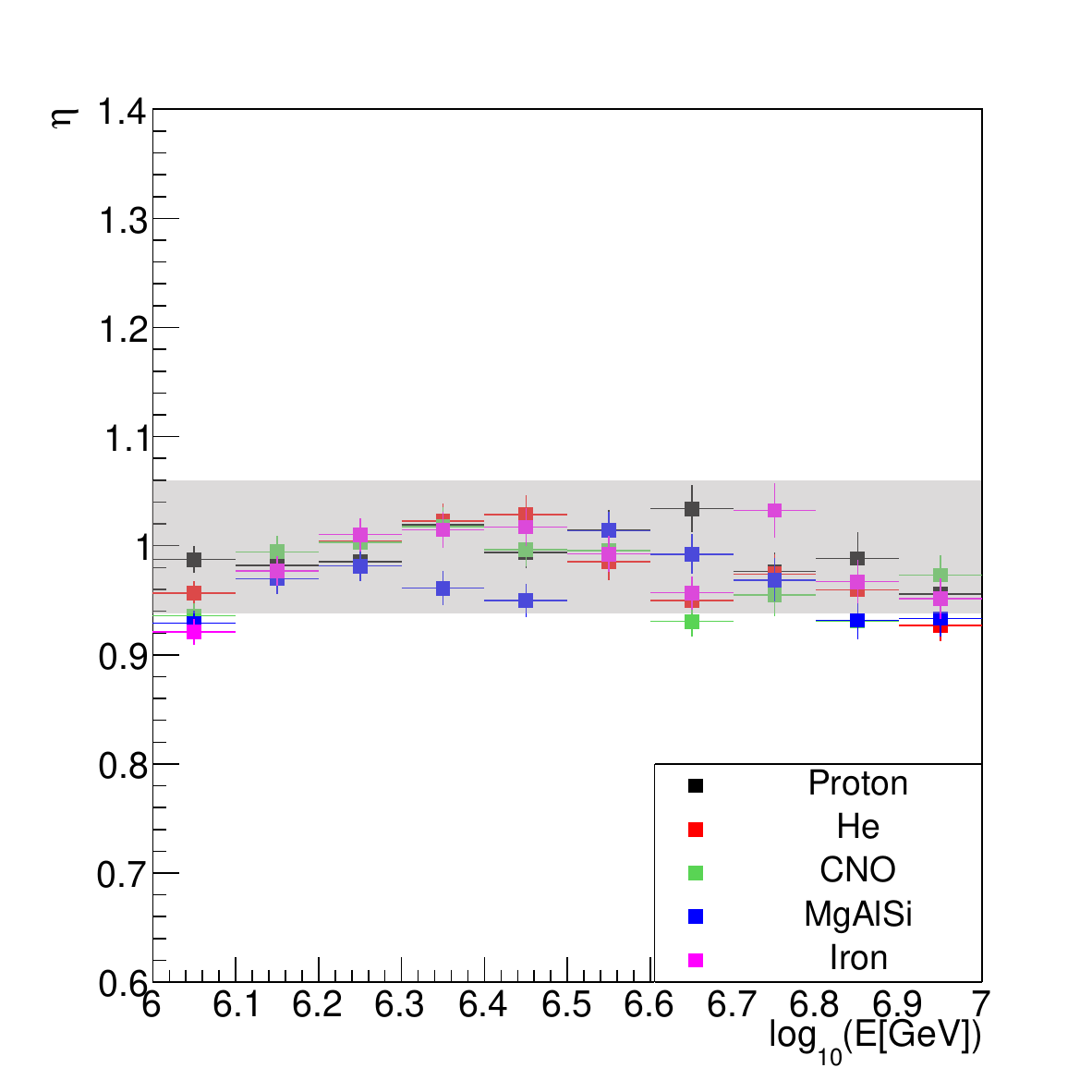}
    \caption{Relation between mean value of $\eta$ and primary energy $E$ for five components. The grey shaded band marks the $\pm6\%$ range.}
    \label{fig:E_eta}
\end{figure}

We validate agreement between fitted and measured thermal neutron densities via the ratio $\eta$ defined in Eq.~\ref{eq:eta}. Here, $\rho_n$ denotes thermal neutron density at $10\,\mathrm{m}$ from the fitted lateral distribution, and $\rho_n^\mathrm{meas}$ the measured density at $10\,\mathrm{m}$. The latter is calculated by dividing the thermal neutron count recorded by EN-detectors within the annulus around $R_{p}=10\,\mathrm{m}$ by the annular geometric area.
\begin{equation}
\label{eq:eta}
\eta=\frac{\rho_{n}}{\rho_{n}^{meas}}
\end{equation}
The $\eta$ distribution for proton showers is shown in Fig.~\ref{fig:proton-etanu} as an example. The relations between the mean value of $\eta$ and the primary energy $E$ for different components are shown in Fig.~\ref{fig:E_eta}. As seen, the mean offset between fitted and measured quantities for different primary compositions and energies lies mostly within $\pm6\%$, as highlighted by the grey shaded band in Fig.~\ref{fig:E_eta}.

% 插入表格
%\begin{table}
%\centering
%\renewcommand{\arraystretch}{1.5} % 调整行高
%\caption{Mean value and RMS of thermal neutron $\eta$ distribution of the five components. }
%\begin{tabular}{|c|c|c|c|c|c|}
%\hline

\begin{figure}[htbp]
    \centering
    \includegraphics[width=0.48\textwidth]{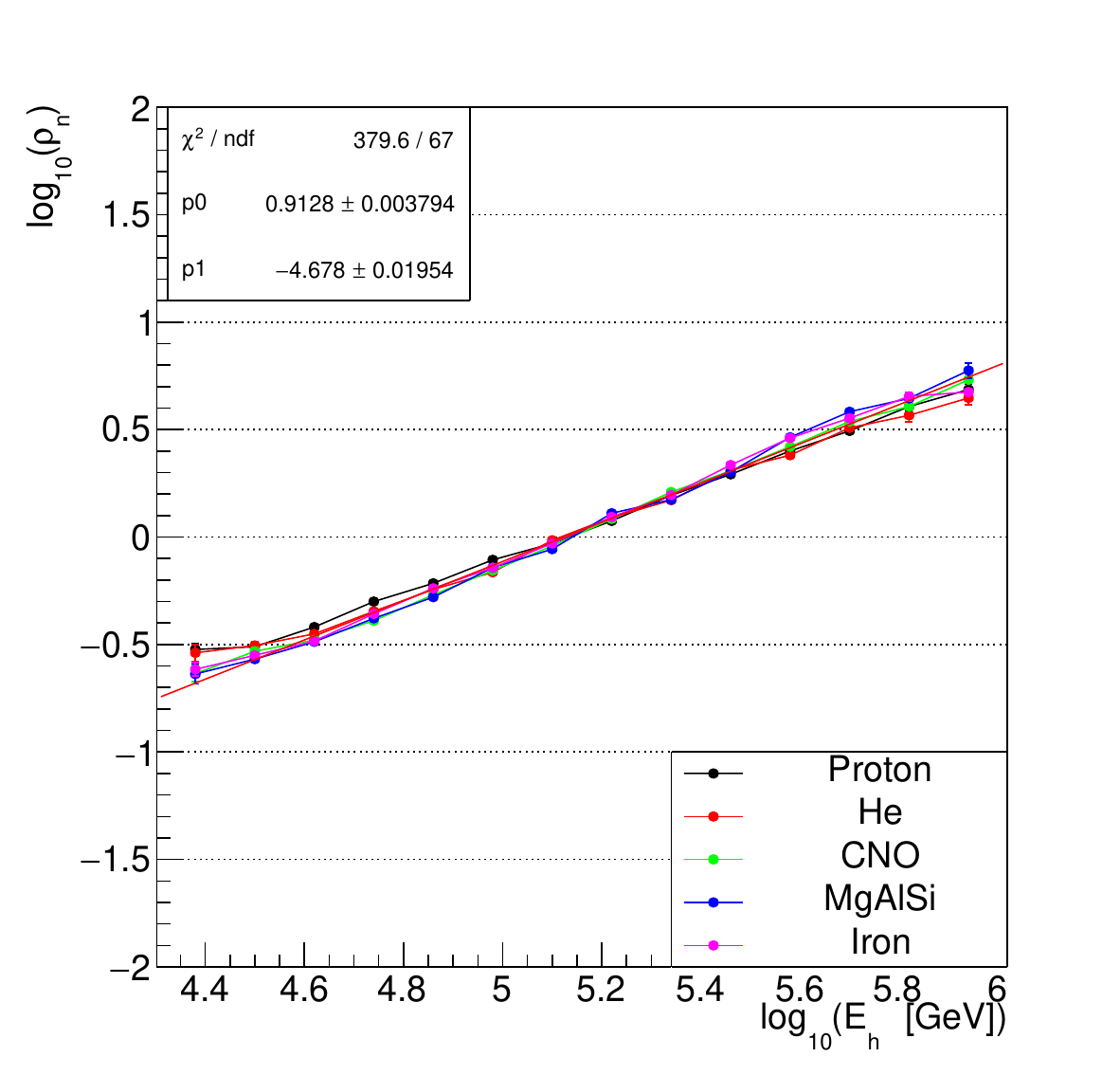}
    \includegraphics[width=0.48\textwidth]{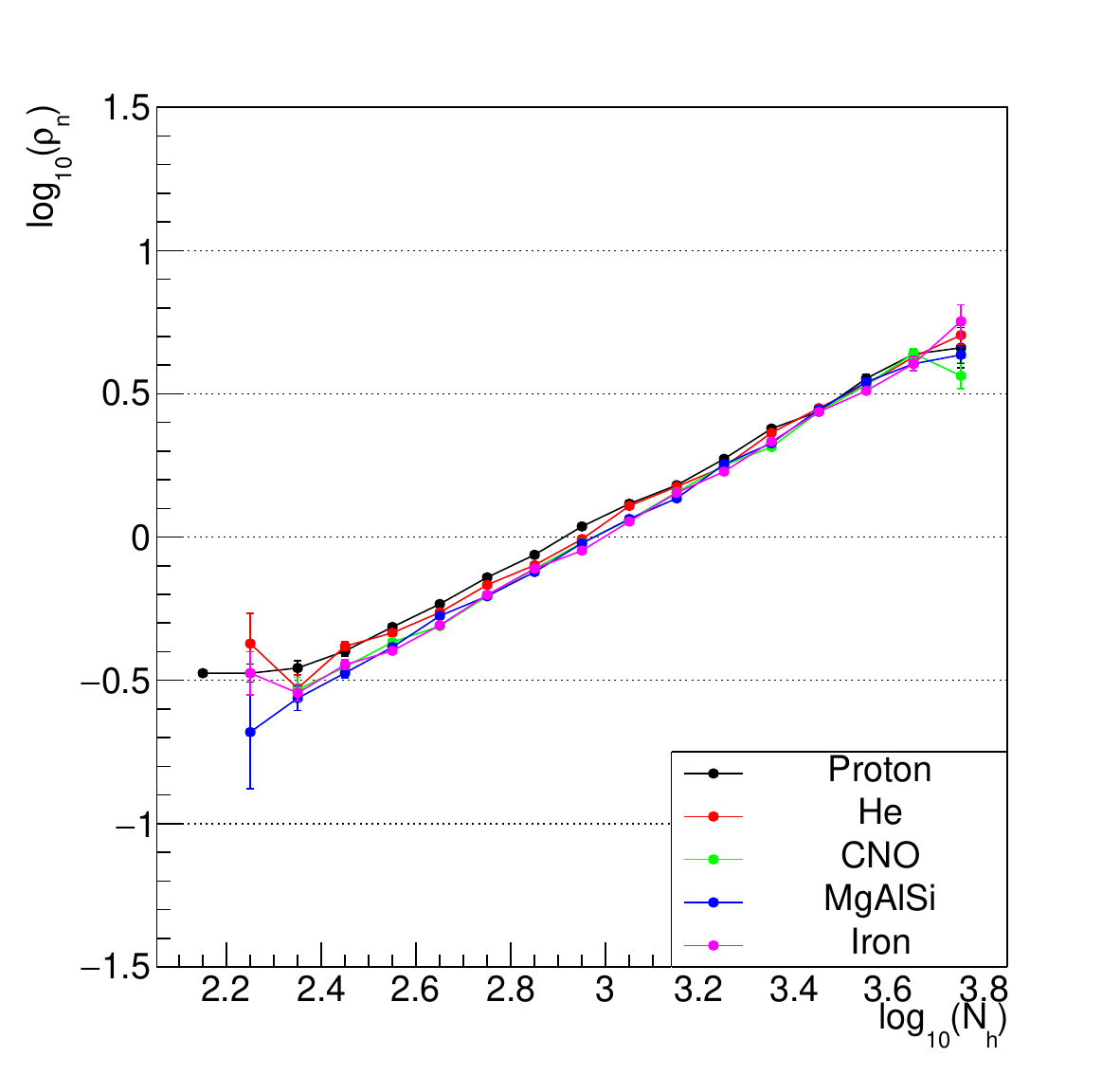}
    \caption{Left: relation between thermal neutron density at 10 m $\rho_n$ and sum of all hadrons energy $E_h$ for different components; Right: relation between thermal neutron density at 10 m $\rho_n$ and the number of high-energy hadrons (>10 GeV, $N_h$) for different components.}
    \label{fig:rhon-hadron}
\end{figure}

Since thermal neutrons are produced by hadrons in EAS, Fig.~\ref{fig:rhon-hadron} further investigates the thermal neutron density $\rho_n$ in relation to secondary hadrons: left panel shows sum of all hadrons energy in EAS, right panel shows the number of high-energy hadrons (above 10 GeV) in EAS. As seen, thermal neutron density $\rho_n$ depends mainly on the sum of all hadrons energy or the number of high-energy hadrons. The number excludes low-energy hadrons which do not contribute to evaporation neutron production in air showers~\cite{Stenkin_2013}. The relationship with sum of all hadron energy (or the number of high-energy hadrons) can be well fitted by a linear function on a log-log scale with a slope of about 0.91. This demonstrates that thermal neutrons are a good proxy for the hadronic component of EAS.

%++++++++++++++++++++++++++++++++++++++++++++++++++++

\section{Summary}
\label{summary}
This study investigates cosmic-ray air shower reconstruction in the knee region using  ENDA simulation data. The lateral distribution functions of thermal neutrons are carefully selected and optimized for the fitting parameters individually. The lateral distribution of thermal neutrons measured by ENDA is analyzed. A comparison of the NKG and double-exponential functions for describing this distribution using MC simulation data shows that the double-exponential function is more accurate. The mean difference between fitted and measured thermal neutron density at 10 m from the shower axis is within $\sim$6\%.\\\indent
The correlation between thermal neutron density at 10 m from the shower axis and the sum of all hadrons energy or the number of high-energy secondary hadrons in EAS was studied. The results show that thermal neutron density $\rho_n$ depends mainly on the sum of all hadrons energy or the number of high-energy hadrons (above 10 GeV). The relationship between $\rho_n$ and sum of all hadron energy (or the number of high-energy secondary hadrons) can be well fitted by a linear function on a log scale. This demonstrates that thermal neutrons are a good proxy for the hadronic component of EAS.\\\indent
The reconstruction results are significant for capabilities of energy determination and composition separation of the array, and then accurate measurement of cosmic ray light component energy spectrum by using ENDA-64 and LHAASO in the near future.

\section*{Acknowledgements}
\hspace{2em}This work was supported in China by National Key R\&D Program of China (grant no. 2025YFB3003602), the National Natural Science Foundation of China (NSFC, No.12320101005, No.12205244, No.12263005 and No.12373105) Natural Science Foundation of the Tibet Autonomous Region (grant No.XZ202501ZR0058). The authors would like to thank the LHAASO collaboration to provide simulated events with CORSIKA and reconstruct shower direction and core position. The authors would like to thank all staff members who work year-round at the LHAASO site, which is located at an altitude of 4,400 meters above sea level, for maintaining the detectors and ensuring the electricity supply and other components of the experiment run smoothly.
\section*{Competing interests}
On behalf of all authors, the corresponding author states that there is no conflict of interest.

%\bibliographystyle{unsrt}
\bibliography{biblio}% common bib file

\end{document}